\documentclass[11pt,epsf]{article}
\usepackage{amsthm,amstext}
\usepackage{amsmath}
\usepackage{graphicx}
\usepackage{amsfonts}
\usepackage{amssymb}
\theoremstyle{plain}

\theoremstyle{definition}

\theoremstyle{remark}

\date{}
\title{\bf  The algebraic hyperstructure of elementary particles in physical theory}\vspace{.25 in}
\author{{\bf A. Dehghan Nezhad}$^a$, {\bf M. Nadjafikhah}$^b$ and {\bf
S.M. Moosavi Nejad}$^c$ \\ \\
$^a$Faculty of Mathematics, Yazd University, Yazd, Iran\\
{\tt  anezhad@yazduni.ac.ir}\\
$^b$School of Mathematics, Iran University of Science and Technology,\\ Narmak, Tehran, Iran\\
{\tt m\_nadjafikhah@iust.ac.ir}\\
$^c$Faculty of Physics, Yazd University, Yazd, Iran\\ and
School of particles and accelerators, Institute for research in \\ fundamental science (IPM), P.O.Box 19395-5531, Tehran, Iran\\
 {\tt  mmoosavi@yazduni.ac.ir}}

\begin{document}
\def\di{\displaystyle}
\maketitle
\begin{abstract}
Algebraic hyperstructures represent a natural extension of
classical algebraic structures. In a classical algebraic
structure, the composition of two elements is an element, while in
an algebraic hyperstructure, the composition of two elements is a
set. Algebraic hyperstructure theory has a multiplicity of
applications to other disciplines. The main purpose of this paper
is to provide examples of hyperstructures associated with
elementary particles in physical theory.
\end{abstract}

\section{Introduction}
Algebraic hyperstructures represent a natural extension of
classical algebraic structures and they were introduced in 1934 by
the French mathematician F.~Marty \cite{M}. In a classical
algebraic structure, the composition of two elements is an
element, while in an algebraic hyperstructure, the composition of
two elements is a set. Since then, hundreds of papers and several
books have been written on this topic. One of the first books,
dedicated especially to hypergroups, is ``Prolegomena of
Hypergroup Theory", written by P. Corsini in 1993 \cite{C1}.
Another book on ``Hyperstructures and Their Representations", by
T. Vougiouklis, was published one year later \cite{V}.

On the other hand, algebraic hyperstructure theory has a
multiplicity of applications to other disciplines: geometry,
graphs and hypergraphs, binary relations, lattices, groups, fuzzy
sets and rough sets, automata, cryptography, codes, median
algebras, relation algebras, $C^*$-algebras, artificial
intelligence, probabilities and so on. A recent book on these
topics is ``Applications of Hyperstructure Theory", by P.Corsini
and V. Leoreanu, published by Kluwer Academic Publishers in 2003
\cite{CL}. We mention here another important book for the
applications in Geometry and for the clearness of the exposition,
written by W. Prenowitz and J. Jantosciak \cite{PJ}. Another
monograph is devoted especially to the study of Hyperring Theory,
written by Davvaz and Leoreanu-Fotea \cite{DL}. It begins with
some basic results concerning ring theory and algebraic
hyperstructures, which represent the most general algebraic
context, in which the reality can be modelled. Several kinds of
hyperrings are introduced and analyzed in this book. The volume
ends with an outline of applications in Chemistry and Physics
\cite{DD1, DD2}, canalizing several special kinds of
hyperstructures: $e-$hyperstructures and transposition
hypergroups. The theory of suitable modified hyperstructures can
serve as a mathematical background in the field of quantum
communication systems.

The main purpose of this paper is to provide examples of
hyperstructures associated with elementary particles in physical
theory.\\
\section{General n-ary hyperstuctures }
Throughout this paper, the symbol $H_1, H_2, \ldots , H_n$ will
denote $n$ nonempty sets. Where $P^\ast (\bigcup_{i=1}^n H_i)$
denotes the set of all non empty
subsets of $\bigcup_{i=1}^n H_i$.\\

\paragraph{Definition 2.1.}{\it A general $n-$ary hyperstructure is $n$ non
empty sets $H_1, H_2, \ldots , H_n$ together with a
hyperoperation,
$$f : H_1\times H_2 \times \ldots \times H_n \longrightarrow P^\ast ( \bigcup_{i=1}^n H_i)$$
$$(x_1, \ldots \, x_n) \mapsto f(x_1, \ldots \, x_n) \subseteq (\bigcup_{i=1}^n H_i)-\emptyset.$$}

\indent  Let $f$ be a general $n$-ary hyperoperation on $H_1, H_2,
\ldots, H_n$ and $A_i,$ subsets of $H_i$ for all $i=1,\ldots ,n $.
We define
\[
f(A_1, \ldots , A_n)=\bigcup \{f(x_1, \ldots , x_n)| \ x_i \in
A_i, i=1, \ldots ,n \}.
\]
\\
 We denote by $H^n$ the cartesian product
$H\times \ldots \times H$ where $H$ appears $n$ times. An element
of $H^n$ will be denoted by $(x_1, \ldots , x_n)$ where $x_i \in
H$ for any $i$ with $1\leq i \leq n$. In general, a mapping $f:H^n
\longrightarrow {\cal P}^*(H)$ is called an {\it $n$-ary
hyperoperation} and $n$ is called the {\it order of
hyperoperation}. A {\it hyperalgebra } $(H, f)$ is a non-empty set
$H$ with one $n$-ary
hyperoperations $f$. \\

\paragraph{Remark 2.2.} A general hyperoperation $\ast : X \times Y
\longrightarrow P^\ast (X \cup Y)$ yields a general hyperoperation
$$\otimes : P^\ast (X) \times P^\ast(Y) \longrightarrow P^\ast (X
\cup Y)$$ defined by $\di A \otimes B =
\bigcup_{a \in A, b \in B} a\ast b $.\\

Conversely a general hyperoperation on $P^\ast (X) \times
P^\ast(Y)$ yields a general hyperoperation on $X \times Y$,
defined
by $x \ast y = \{x\} \otimes \{y\}.$\\
In the above definition if $A \subseteq X, \ B  \subseteq Y, \ x
\in X, \ y \in Y, $ then we define, $$A \ast y = A \ast \{y\} =
\bigcup_{a \in A} a\ast y, \ x \ast B = \{x\} \ast B = \bigcup_{b
\in B} x\ast b, $$
$$A \otimes B  =\bigcup_{a
\in A, b\in B} a\ast b. $$

\paragraph{Remark 2.3.} If we let $X=Y=H$, then we obtain the
hyperstructure theory.\\

\section{Algebraic hyperstructures} In this subsection, we
summarize the preliminary definitions and results required in the
sequel. \\

\paragraph{Definition 3.1.} Let $H$ be a non-empty set and let ${\cal P}^*(H)$ be the set of
all non-empty subsets of $H$.
\begin{itemize}
\item [(i)] A {\it hyperoperation} on $H$ is a
map $\otimes:H\times H\longrightarrow {\cal P}^*(H)$ and the
couple $(H,\otimes)$ is called a {\it hypergroupoid}. If $A$ and
$B$ are non-empty subsets of $H$, then we denote
$$A\otimes B=\bigcup_{a\in A,\,b\in B}a\otimes b,\ \ x\otimes
A=\{x\}\otimes A\mbox{\ \ and\ \ }A\otimes x=A\otimes\{x\}.$$

\item [(ii)] A hypergroupoid  $(H,\otimes)$ is called a {\it
semihypergroup} if for all $x,y,z$ of $H$ we have $(x\otimes
y)\otimes z=x\otimes(y\otimes z)$, which means that
$$\displaystyle \bigcup_{u\in x\otimes y}u\otimes z=\bigcup_{v\in
y\otimes z}x\otimes v.$$

\item [(iii)] We say that a  semihypergroup $(H,\otimes)$ is a {\it
hypergroup} if for all $x\in H$, we have $x\otimes H=H\otimes
x=H.$ A hypergroupoid $(H,\otimes )$ is an {\it $H_v$-group}, if
for all $x,y,z\in H,$ the following conditions hold: \item [(1)]
$x\otimes (y\otimes z) \cap (x\otimes y)\otimes z \not =
\emptyset$ (weak associativity), \item [(2)] $x\otimes H=H\otimes
x=H$ (reproduction).
\end{itemize}


\paragraph{Definition 3.2.} Let $( L, \oplus ),$ be a
 $H_v$-group, and $K$ be a nonempty subset of $L$. Then $K$ is called $H_v$-subgroup of
$(L, \oplus)$ if $ a \oplus b \in {\cal P}^\ast (K)$ for all $a, b
\in K$.
 That is to say, $K$ is an  $H_v$-subgroup of
$(L, \oplus ) $ if and only if $K$ is closed under the binary
hyperoperation on $L$.\\

The concept of $H_v$-structures constitute a
 generalization of the well-known algebraic hyperstructures (
 hypergroup, hyperring, hypermodule, and so on) (for example you see [1]).\\

\section{Physical example (elementary particles)}

Since long time ago, one of the most important questions is that what is our universe made of?.
 To answer this question many efforts have been yet done. Since 1897 when the electron was discovered by
  J. J. Thomson, the elementary particle physics was born. Nowadays, the biggest particle accelerator which
  is called  the Large Hadron Collider  (LHC) at CERN in France is being applied to find the last elementary
   particle(Higgs boson). The existence of this particle is necessary to understand our universe.\\

In particle physics, an elementary particle or fundamental particle is a particle which have no substructure,
 i.e. it is not known to be made up of smaller particles. If an elementary particle truly has no substructure,
  then it is one of the basic building blocks of the universe from which all other particles are made.
  To describe the elementary particles and the interacting forces between them, some different theories
   are proposed that their most important is the Standard Model \cite{Muta}. The Standard Model(SM) of
   elementary particles has proved to be extremely successful during the past three decades. It has shown
    to be a well established theory. All predictions based on the SM have been experimentally verified and
     most of its parameters have been fixed. The only part of the SM that has not been directly experimentally
     verified yet is the Higgs sector. In the SM, the Quarks, Leptons and Gauge bosons are introduced as the
      elementary particles. The SM of particle physics contains six types of quarks, known as flavors: Up,
       Down, Charm, Strange, Bottom and Top plus their corresponding antiparticles.  For every particle there
        is a corresponding type of antiparticle, for every quark it is known as antiquark, that differs from
         the particle only in some of its properties(like electric charge) which have equal magnitude but
          opposite sign. Since the quarks are never found  in isolation, therefore quarks combine to form
          composite particles which are called Hadrons, see Ref.\cite{hadrons}. In particle physics, hadrons
           are composite particles made of quarks which are categorized into two families: baryons (made of
           three quarks) and mesons (made of one quark and one antiquark).\\

In the SM, gauge bosons consist of the photons($\gamma$), gluons($g$), $W^\pm$ and $Z$ bosons act as carriers
 of the fundamental forces of nature\cite{parta}. In fact, interactions between the particles are described by
  the exchange of gauge bosons. The third group of the elementary particles are leptons. There are six type of
   leptons including the electron($e$), electron neutrino($\nu_e$), muon($\mu$), muon neutrino($\nu_\mu$),
   tau($\tau$) and tau neutrino($\nu_\tau$). Every lepton has a corresponding antiparticle-these antiparticles
    are known as antileptons. Leptons are an important part of the SM, especially the electrons which are one
    of the components of atoms. Since the leptons can be found freely in the universe and they are one of the
     important groups of the elementary particles, in this article we only concentrate on this group of
      particles.

\subsection{Leptons}

In the Standard Model, leptons still appear to be structureless. In this model, there are six flavors of
leptons and six corresponding antiparticles. They form three generations \cite{parta,partb}. The first
generation is the electronic leptons, comprising the electron($e$), electron neutrino($\nu_e$) and their
corresponding antiparticles, i.e. positron($e^+$) and electron antineutrino ($\overline{\nu}_e$).
 The second generation is the muonic leptons, including muon($\mu$), muon neutrino($\nu_\mu$),
  antimuon($\mu^+$) and muon antineutrino($\overline{\nu}_\mu$). The third is the tauonic leptons,
  consist of tau($\tau$), tau neutrino($\nu_\tau$), antitau($\tau^+$) and tau antineutrino
  $(\overline{\nu}_\tau)$. Therefore,  there are 12 particles $\{e, \nu_e, \mu, \nu_\mu, \tau,
   \nu_\tau, e^+, \overline{\nu}_e, \mu^+, \overline{\nu}_\mu, \tau^+, \overline{\nu}_\tau\}$
   in the leptons group. In the leptons group, the electron, muon and tau have the electric
   charge $Q=-1$ (the charge of a particle is expressed in unit of the electron charge) and the
    neutrinos are neutral. According to the definition of antiparticle, the electric charge of positron,
     antimuon and antitau is $Q=+1$ but the antineutrinos are neutral as well as neutrinos.
     The main difference between the neutrinos and antineutrinos is in the other quantum numbers such
      as leptonic numbers, see Refs.~\cite{parta,partb}. In the SM, leptonic numbers are assigned  to
      the members of every generation of leptons.  Electron and electron neutrino have an electronic
      number of $Le = 1$  while muon and muon neutrino have a muonic number of $L_\mu = 1$ and tau and
      tau neutrino have a tauonic number of $L_\tau = 1$. The antileptons have their respective generation's
       leptonic numbers of $-1$. These numbers are classified in Table
       \ref{a1}.
\\

\begin{table}[h]
\begin{center}
\begin{tabular}{|c|cccc|cccc|cccc|}
\hline Classify & First Generation\hspace{-3cm} &&&& Second
Generation\hspace{-3cm} &&&& Third Generation\hspace{-3cm} \mbox{
} &&&
\\ \hline\
Leptons  & $e$    & $\nu_e$  &  $e^+$  &  $\overline{\nu}_e$  & $\mu$ &  $\nu_\mu$ &  $\mu^+$ &  $\overline{\nu}_\mu$  & $\tau$ & $\nu_\tau$ & $\tau^+$ &  $\overline{\nu}_\tau$\\
\hline\
$Q$         & $-1$   &  $0$     &  $+1$   &        $0$           & $-1$  &   $0$      &   $+1$   &     $0$                &   $-1$ &   $0$      &   $+1$   &    $0$               \\
\hline
$L_e$       & $1$    &  $1$     &  $-1$   &        $-1$          & $0$  &   $0$       &   $0$    &     $0$                &   $0$ &    $0$      &   $0$    &    $0$               \\
\hline
$L_\mu$     & $0$    &  $0$     &  $0$   &        $0$            & $1$  &   $1$       &   $-1$   &     $-1$               &   $0$ &   $0$      &   $0$   &    $0$               \\
\hline
$L_\tau$    & $0$    &  $0$     &  $0$   &        $0$           & $0$  &   $0$        &   $0$    &     $0$                &   $1$ &   $1$      &   $-1$   &    $-1$               \\
\hline
\end{tabular}
\caption{Leptons classification to three generations. $Q$ stands for charge in unit of the electron charge.
$L_e, L_\mu$ and $L_\tau$ stand for the electronic, muonic and tauonic numbers, respectively. \label{a1}}
\end{center}
\end{table}
 In every interaction the leptonic numbers are conserved.
Conservation of the leptonic numbers means that
 the number of leptons of the same type remains the same when particles interact. This implies that leptons
  and antileptons must be created in pairs of a single generation. For example, the following processes are
   allowed under conservation of the leptonic numbers:
$$
e+\nu_e \rightarrow \{ e+\nu_e\}=\{ e,\nu_e\} \ or \ \mu+\nu_\mu
\rightarrow \{\mu+\nu_\mu\}=\{\mu,\nu_\mu\},
$$
 where in the first
interaction, the  electronic numbers and in the second interaction
the munic numbers are
 conserved. In other interactions, outgoing particles might be different, therefore all leptonic numbers must
  be checked. In the following interactions both the electronic and muonic numbers are conserved:
$$
e+\nu_\mu \rightarrow \{ e+\nu_\mu \ or \ \mu+\nu_e \}=\{
e,\nu_\mu, \mu,\nu_e \}.
$$
 The only conditions required to occur a leptonic interaction are
the conservation of the electric charges and the leptonic numbers.
Conservation of the leptonic numbers  means that the electronic,
muonic and tauonic
 numbers must be conserved sepreately.  Considering these conservation rules, for the electron-positron
  interaction the interacting modes are:
$$
e+e^+ \rightarrow \{e+e^+\ or \ \mu+\mu^+\ \ or \ \tau+\tau^+\ or\
\nu_e+\overline{\nu}_e\ or \ \nu_\mu+\overline{\nu}_\mu\ or \
\nu_\tau+\overline{\nu}_\tau \}=L.
$$

 Other interactions between the members of the leptons group are
shown in Table 2.
 To arrange this table we avoided writing the repeated symbols. For example in the productions
  of the electron-electron interaction  we only write $e$ instead of $e+e$. All the interactions
  shown in Table 2 are in the first order. It means in the higher order other particles can
   be produced that we do not consider them. For example in the electron-electron scattering
   (Muller scattering) one or several photons might be appeared in productions of the interaction,
    i.e. $e+e\rightarrow e+e+\gamma$ or in the electron-positron scattering(Bhabha scattering \cite{bhabha})
     we can have: $e+e^+ \rightarrow e+e^++\gamma$. There also exist other processes that we do not consider
      them in this work. For example:  $e+e^+\rightarrow \gamma+\gamma$,  $e+e^+\rightarrow W^-+W^+$,
       $\tau+\tau^+\rightarrow Z^0+Z^0$ and so on.\\

\section {The algebraic hyperstructure of leptons}

Contemporary investigations of hyperstructures and their
applications yield many relationships and connections between
various fields of mathematics.\\

Besides the motivation for investigation of hyperstructures coming
from noncommutative algebra, geometrical structures and other
mathematical fields there exist such physical fenomena as the
nuclear fission. Nuclear fission occurs when a heavy nucleus, such
as U23 5 , splits, or fissions, into two smaller nuclei. As a
result of this fission process we can get several dozens of
diffierent combinations of two medium-mass elements and several
neutrons (as barium $Ba^{141}$ and krypton $Kr^{92}$ and 3
neutrons; strontium $Sr^{94}$ , xenon $Xe^{14}$ 0 and 2 neutrons;
lanthanum $La^{147}$ , bromum $Br^{87}$ and 2 neutrons; $Sn^{132}$
, $Mo^{101}$ and 3 neutrons and so on) . More precisely, the input
of this reaction is always the same|the heavy uranium is bombarded
with neutrons, but the result is in general different|there are
about 90 different daughter nuclei that can be formed. The fission
also results in the production of several neutrons, typically two
or three. On the average about 2.5 neutrons are released per
event. In any fission equation, there are many combinations of
fission fragments, but they always satisfy the requirements of
conservation of energy and charge.\\

Another typical example of the situation when the result of
interaction between two particles is the whole set of particles is
the interaction between a foton with certain energy and an
electron. The result of this interaction is not deterministic. A
photo-electric effect or Coulomb repulsion effect or changeover of
foton onto a pair electron and positron can arise.\\

It is to be noted that a similar situation which occurs during
uranium fission appears during several nuclear fission, too. The
result depends on conditions. Although the input 2 particles are
the same, the output can be variant. It can differ both in the
number of arising particles and in their kind.\\

Another motivation for investigation of hyperstructures yields
from technical processes as a time sequence of military car
repairs with respect to its roadability consequences and its
operational behavior. In this section we describe a certain
construction  of hyperstructures belonging to the important class
of elementary particles. We find the algebraic hyperstructure of
leptons.
\begin{table} \tiny

$$
\begin{array}{|c|c|c|c|c|c|c|c|c|c|c|c|c|} \hline\

\otimes\!\!&\!\!\di e\!\!&\!\!\di \nu_e\!\!&\!\!\di
e^+\!\!&\!\!\di \bar\nu_e\!\!&\!\!\di \mu\!\!&\!\!\di
\nu_\mu\!\!&\!\!\di \mu^+\!\!&\!\!\di \bar\nu_\mu\!\!&\!\!\di
\tau\!\!&\!\!\di \nu_\tau\!\!&\!\!\di \tau^+\!\!&\!\!\di
\bar\nu_\tau \\ \hline
e\!\!&\!\!\di e\!\!&\!\!\di e\atop\di \nu_e\!\!&\!\!\di
L\!\!&\!\!\di
\begin{array}{c} \di e\\ \di \mu\\ \di \tau\\ \di \bar\nu_e\\ \di \bar\nu_\mu\\ \di \bar\nu_\tau\end{array}
\!\!&\!\!\di e\atop\di \mu\!\!&\!\!\di \begin{array}{c} \di e\\
\di \mu\\ \di \nu_e\\ \di \nu_\mu
\end{array}\!\!&\!\!\di \begin{array}{c} \di e\\ \di \mu^+\\ \di \bar\nu_\mu\\ \di \nu_e\end{array}
\!\!&\!\!\di e\atop\di \bar\nu_\mu\!\!&\!\!\di e\atop\di
\tau\!\!&\!\!\di \begin{array}{c}
\di e\\ \di \tau\\ \di \nu_e\\ \di \nu_\tau\end{array}\!\!&\!\!\di \begin{array}{c} \di e\\ \di \tau^+\\
\bar\nu_\tau\\ \di \nu_e\end{array}\!\!&\!\!\di e\atop\di \bar\nu_\tau\\[1mm]\hline\
\nu_e\!\!&\!\!\di e\atop\di \nu_e\!\!&\!\!\di \nu_e\!\!&\!\!\di \begin{array}{c} \di e^+\\ \di \mu^+\\ \di \tau^+\\ \di \nu_e\\ \di \nu_\mu\\ \di \nu_\tau\end{array}\!\!&\!\!\di L\!\!&\!\!\di \begin{array}{c} \di e\\ \di \mu\\ \di \nu_e\\ \di \nu_\mu\end{array}\!\!&\!\!\di \nu_e\atop\di \nu_\mu\!\!&\!\!\di \mu^+\atop\di \nu_e\!\!&\!\!\di \begin{array}{c} \di e\\ \di \mu^+\\ \di \bar\nu_\mu\\ \di \nu_e\end{array}\!\!&\!\!\di \begin{array}{c} \di e\\ \di \tau\\ \di \nu_e\\ \di \nu_\tau\end{array}\!\!&\!\!\di \nu_e\atop\di \nu_\tau\!\!&\!\!\di \tau^+\atop\di \nu_e\!\!&\!\!\di \begin{array}{c} \di e\\ \di \tau^+\\ \di \bar\nu_\tau\\ \di \nu_e\end{array}\\[1mm]\hline\
e^+\!\!&\!\!\di L\!\!&\!\!\di \begin{array}{c} \di e^+\\ \di \mu^+\\ \di \tau^+\\ \di \nu_e\\ \di \nu_\mu\\ \di \nu_\tau\end{array}\!\!&\!\!\di e^+\!\!&\!\!\di e^+\atop\di \bar\nu_e\!\!&\!\!\di \begin{array}{c} \di e^+\\ \di \mu\\ \di \bar\nu_e\\ \di \nu_\mu\end{array}\!\!&\!\!\di e^+\atop\di \nu_\mu\!\!&\!\!\di e^+\atop\di \mu^+\!\!&\!\!\di \begin{array}{c} \di e^+\\ \di \mu^+\\ \di \bar\nu_e\\ \di \bar\nu_\mu\end{array}\!\!&\!\!\di \begin{array}{c} \di e^+\\ \di \tau\\ \di \bar\nu_e\\ \di \nu_\tau\end{array}\!\!&\!\!\di e^+\atop\di \nu_\tau\!\!&\!\!\di e^+\atop\di \tau^+\!\!&\!\!\di \begin{array}{c} \di e^+\\ \di \tau^+\\ \di \bar\nu_e\\ \di \bar\nu_\tau\end{array}\\[1mm]\hline\
\bar\nu_e\!\!&\!\!\di \begin{array}{c} \di e\\ \di \mu\\ \di \tau\\ \di \bar\nu_e\\ \di \bar\nu_\mu\\ \di \bar\nu_\tau\end{array}\!\!&\!\!\di L\!\!&\!\!\di e^+\atop\di \bar\nu_\mu\!\!&\!\!\di \bar\nu_e\!\!&\!\!\di \mu\atop\di \bar\nu_e\!\!&\!\!\di \begin{array}{c} \di e^+\\ \di \mu\\ \di \bar\nu_e\\ \di \nu_\mu\end{array}\!\!&\!\!\di \begin{array}{c} \di e^+\\ \di \mu^+\\ \di \bar\nu_e\\ \di \bar\nu_\mu\end{array}\!\!&\!\!\di \bar\nu_e\atop\di \bar\nu_\mu\!\!&\!\!\di \tau\atop\di \bar\nu_e\!\!&\!\!\di \begin{array}{c} \di e^+\\ \di \tau\\ \di \bar\nu_e\\ \di \nu_\tau\end{array}\!\!&\!\!\di \begin{array}{c} \di e^+\\ \di \tau^+\\ \di \bar\nu_e\\ \di \bar\nu_\tau\end{array}\!\!&\!\!\di \bar\nu_e\atop\di \bar\nu_\tau\\[1mm]\hline\
\mu\!\!&\!\!\di e\atop\di \mu\!\!&\!\!\di \begin{array}{c} \di e\\ \di \mu\\ \di \nu_e\\ \di \nu_\mu\end{array}\!\!&\!\!\di \begin{array}{c} \di e^+\\ \di \mu\\ \di \bar\nu_e\\ \di \nu_\mu\end{array}\!\!&\!\!\di \mu\atop\di \bar\nu_e\!\!&\!\!\di \mu\!\!&\!\!\di \mu\atop\di \nu_\mu\!\!&\!\!\di L\!\!&\!\!\di \begin{array}{c} \di e\\ \di \mu\\ \di \tau\\ \di \bar\nu_e\\ \di \bar\nu_\mu\\ \di \bar\nu_\tau\end{array}\!\!&\!\!\di \mu\atop\di \tau\!\!&\!\!\di \begin{array}{c} \di \mu\\ \di \tau\\ \di \nu_\mu\\ \di \nu_\tau\end{array}\!\!&\!\!\di \begin{array}{c} \di \mu\\ \di \tau^+\\ \di \bar\nu_\tau\\ \di \nu_\mu\end{array}\!\!&\!\!\di \mu\atop\di \bar\nu_\tau\\[1mm]\hline\
\nu_\mu\!\!&\!\!\di \begin{array}{c} \di e\\ \di \mu\\ \di \nu_e\\ \di \nu_\mu\end{array}\!\!&\!\!\di \nu_e\atop\di \nu_\mu\!\!&\!\!\di e^+\atop\di \nu_\mu\!\!&\!\!\di \begin{array}{c} \di e^+\\ \di \mu\\ \di \bar\nu_e\\ \di \nu_\mu\end{array}\!\!&\!\!\di \mu\atop\di \nu_\mu\!\!&\!\!\di \nu_\mu\!\!&\!\!\di \begin{array}{c} \di e^+\\ \di \mu^+\\ \di \tau^+\\ \di \nu_e\\ \di \nu_\mu\\ \di \nu_\tau\end{array}\!\!&\!\!\di L\!\!&\!\!\di \begin{array}{c} \di \mu\\ \di \tau\\ \di \nu_\mu\\ \di \nu_\tau\end{array}\!\!&\!\!\di \nu_\mu\atop\di \nu_\tau\!\!&\!\!\di \tau^+\atop\di \nu_\mu\!\!&\!\!\di \begin{array}{c} \di \mu\\ \di \tau^+\\ \di \bar\nu_\tau\\ \di \nu_\mu\end{array}\\[1mm]\hline\
\mu^+\!\!&\!\!\di \begin{array}{c} \di e\\ \di \mu^+\\ \di \bar\nu_\mu\\ \di \nu_e\end{array}\!\!&\!\!\di \mu^+\atop\di \nu_e\!\!&\!\!\di e^+\atop\di \mu^+\!\!&\!\!\di \begin{array}{c} \di e^+\\ \di \mu^+\\ \di \bar\nu_e\\ \di \bar\nu_\mu\end{array}\!\!&\!\!\di L\!\!&\!\!\di \begin{array}{c} \di e^+\\ \di \mu^+\\ \di \tau^+\\ \di \nu_e\\ \di \nu_\mu\\ \di \nu_\tau\end{array}\!\!&\!\!\di \mu^+\!\!&\!\!\di \mu^+\atop\di \bar\nu_\mu\!\!&\!\!\di \begin{array}{c} \di \mu^+\\ \di \tau\\ \di \bar\nu_\mu\\ \di \nu_\tau\end{array}\!\!&\!\!\di \mu^+\atop\di \nu_\tau\!\!&\!\!\di \mu^+\atop\di \tau^+\!\!&\!\!\di \begin{array}{c} \di \mu^+\\ \di \tau^+\\ \di \bar\nu_\mu\\ \di \bar\nu_\tau\end{array}\\[1mm]\hline\
\bar\nu_\mu\!\!&\!\!\di e\atop\di \bar\nu_\mu\!\!&\!\!\di \begin{array}{c} \di e\\ \di \mu^+\\ \di \bar\nu_\mu\\ \di \nu_e\end{array}\!\!&\!\!\di \begin{array}{c} \di e^+\\ \di \mu^+\\ \di \bar\nu_e\\ \di \bar\nu_\mu\end{array}\!\!&\!\!\di \bar\nu_e\atop\di \bar\nu_\mu\!\!&\!\!\di \begin{array}{c} \di e\\ \di \mu\\ \di \tau\\ \di \bar\nu_e\\ \di \bar\nu_\mu\\ \di \bar\nu_\tau\end{array}\!\!&\!\!\di L\!\!&\!\!\di \bar\nu_\mu\mu^+\!\!&\!\!\di \bar\nu_\mu\!\!&\!\!\di \tau\bar\nu_\mu\!\!&\!\!\di \begin{array}{c} \di \mu^+\\ \di \tau\\ \di \bar\nu_\mu\\ \di \nu_\tau\end{array}\!\!&\!\!\di \begin{array}{c} \di \mu^+\\ \di \tau^+\\ \di \bar\nu_\mu\\ \di \bar\nu_\tau\end{array}\!\!&\!\!\di \bar\nu_\mu\atop\di \bar\nu_\tau\\[1mm]\hline\
\tau\!\!&\!\!\di e\atop\di \tau\!\!&\!\!\di \begin{array}{c} \di e\\ \di \tau\\ \di \nu_e\\ \di \nu_\tau\end{array}\!\!&\!\!\di \begin{array}{c} \di e^+\\ \di \tau\\ \di \bar\nu_e\\ \di \nu_\tau\end{array}\!\!&\!\!\di \tau\atop\di \bar\nu_e\!\!&\!\!\di \mu\atop\di \tau\!\!&\!\!\di \begin{array}{c} \di \mu\\ \di \tau\\ \di \nu_\mu\\ \di \nu_\tau\end{array}\!\!&\!\!\di \begin{array}{c} \di \mu^+\\ \di \tau\\ \di \bar\nu_\mu\\ \di \nu_\tau\end{array}\!\!&\!\!\di \tau\atop\di \bar\nu_\mu\!\!&\!\!\di \tau\!\!&\!\!\di \tau\atop\di \nu_\tau\!\!&\!\!\di L\!\!&\!\!\di \begin{array}{c} \di e\\ \di \mu\\ \di \tau\\ \di \bar\nu_e\\ \di \bar\nu_\mu\\ \di \bar\nu_\tau\end{array}\\[1mm]\hline\
\nu_\tau\!\!&\!\!\di \begin{array}{c} \di e\\ \di \tau\\ \di \nu_e\\ \di \nu_\tau\end{array}\!\!&\!\!\di \nu_e\atop\di \nu_\tau\!\!&\!\!\di e^+\atop\di \nu_\tau\!\!&\!\!\di \begin{array}{c} \di e^+\\ \di \tau\\ \di \bar\nu_e\\ \di \nu_\tau\end{array}\!\!&\!\!\di \begin{array}{c} \di \mu\\ \di \tau\\ \di \nu_\mu\\ \di \nu_\tau\end{array}\!\!&\!\!\di \nu_\mu\atop\di \nu_\tau\!\!&\!\!\di \mu^+\atop\di \nu_\tau\!\!&\!\!\di \begin{array}{c} \di \mu^+\\ \di \tau\\ \di \bar\nu_\mu\\ \di \nu_\tau\end{array}\!\!&\!\!\di \tau\atop\di \nu_\tau\!\!&\!\!\di \nu_\tau\!\!&\!\!\di \begin{array}{c} \di e^+\\ \di \mu^+\\ \di \tau^+\\ \di \nu_e\atop\di \\ \di \nu_\mu\\ \di \nu_\tau\end{array}\!\!&\!\!\di L\\[1mm]\hline\
\tau^+\!\!&\!\!\di \begin{array}{c} \di e\\ \di \tau^+\\ \di \bar\nu_\tau\\ \di \nu_e\end{array}\!\!&\!\!\di \tau^+\atop\di \nu_e\!\!&\!\!\di e^+\atop\di \tau^+\!\!&\!\!\di \begin{array}{c} \di e^+\\ \di \tau^+\\ \di \bar\nu_e\\ \di \bar\nu_\tau\end{array}\!\!&\!\!\di \begin{array}{c} \di \mu\\ \di \tau^+\\ \di \bar\nu_\tau\\ \di \nu_\mu\end{array}\!\!&\!\!\di \tau^+\atop\di \nu_\mu\!\!&\!\!\di \mu^+\atop\di \tau^+\!\!&\!\!\di \begin{array}{c} \di \mu^+\\ \di \tau^+\\ \di \bar\nu_\mu\\ \di \bar\nu_\tau\end{array}\!\!&\!\!\di L\!\!&\!\!\di \begin{array}{c} \di e^+\\ \di \mu^+\\ \di \tau^+\\ \di \di \nu_e \\ \nu_\mu\\ \di \nu_\tau\end{array}\!\!&\!\!\di \tau^+\!\!&\!\!\di \tau^+\atop\di \bar\nu_\tau\\[1mm]\hline\
\bar\nu_\tau\!\!&\!\!\di e\atop\di \bar\nu_\tau\!\!&\!\!\di \begin{array}{c} \di e\\ \di \tau^+\\ \di \bar\nu_\tau\\ \di \nu_e\end{array}\!\!&\!\!\di \begin{array}{c} \di e^+\\ \di \tau^+\\ \di \bar\nu_e\\ \di \bar\nu_\tau\end{array}\!\!&\!\!\di \bar\nu_e\atop\di \bar\nu_\tau\!\!&\!\!\di \mu\atop\di \bar\nu_\tau\!\!&\!\!\di \begin{array}{c} \di \mu\\ \di \tau^+\\ \di \bar\nu_\tau\\ \di \nu_\mu\end{array}\!\!&\!\!\di \begin{array}{c} \di \mu^+\\ \di \tau^+\\ \di \bar\nu_\mu\\ \di \bar\nu_\tau\end{array}\!\!&\!\!\di \bar\nu_\mu\atop\di \bar\nu_\tau\!\!&\!\!\di \begin{array}{c} \di e\\ \di \mu\\ \di \tau\\ \di \bar\nu_e\\ \di \bar\nu_\mu\\ \di \bar\nu_\tau\end{array}\!\!&\!\!\di L\!\!&\!\!\di \tau^+\atop\di \bar\nu_\tau\!\!&\!\!\di \bar\nu_\tau\\
\hline
\end{array}
$$
\caption{Interaction between leptons are shown.}

 \end{table}

\paragraph{Theorem 5.1.}
{\it Let $L=\{ e,  \nu_e, e^+, \bar\nu_e, \mu, \nu_\mu, \mu^+,
\bar\nu_\mu, \tau, \nu_\tau, \tau^+, \bar\nu_\tau\}$. Then
$(L,\otimes )$ is an abelian $H_v$-group.}\\

We summarize some results on associativity, in following lemmas.\\

\paragraph{Lemma 5.2.}{\it The list of $[A,B,C]$'s that
$[A,[B,C]]\not\subseteq[[A,B],C]$ is:\\

$[e^+,\mu,e^+]$, $[e^+,\tau,e^+]$, $[e^+,\bar\nu_\tau,e^+]$,
$[\mu,e^+,e^+]$, $[\mu^+,\bar\nu_e,e^+]$, $[\tau,e^+,e^+]$,
$[\tau^+,\bar\nu_e,e^+]$, \\ $[\bar\nu_e,e^+,e^+]$,
$[\bar\nu_e,e^+,\mu]$, $[\bar\nu_e,e^+,\tau^+]$,
$[\bar\nu_e,e^+,\nu_\mu]$, $[\bar\nu_e,e^+,\nu_\tau]$,
$[\bar\nu_\tau,e^+,e^+]$, $[\nu_e,\bar\nu_e,e^+]$,\\
$[\nu_\tau,\bar\nu_e,e^+]$}

\paragraph{Lemma 5.3.} {\it The list of $[A,B,C]$'s that
$[[A,B],C]\not\subseteq[A,[B,C]]$ is:\\

$[e^+,\bar\nu_e,e^+]$, $[\mu,\bar\nu_e,e^+]$,
$[\tau^+,\bar\nu_e,e^+]$, $[\bar\nu_e,e^+,\mu^+]$,
$[\bar\nu_e,e^+,\tau^+]$, $[\bar\nu_e,e^+,\nu_e]$,
$[\bar\nu_e,e^+,\nu_\tau]$,\\ $[\bar\nu_e,\tau^+,\bar\nu_e]$,
$[\bar\nu_e,\bar\nu_e,e^+]$, $[\bar\nu_e,\bar\nu_e,\tau^+]$,
$[\bar\nu_e,\bar\nu_e,\nu_\mu]$, $[\bar\nu_e,\bar\nu_e,\nu_\tau]$,
$[\bar\nu_e,\nu_\mu,\bar\nu_e]$,
$[\bar\nu_e,\nu_\tau,\bar\nu_e]$,\\ $[\nu_\mu,\bar\nu_e,e^+]$,
$[\nu_\tau,\bar\nu_e,e^+]$.}

\paragraph{Lemma 5.4.}
{\it The list of $[A,B,C]$'s that  $[[A,B],C]\neq[A,[B,C]]$ is :\\

$[e^+,\mu,e^+]$, $[e^+,\tau,e^+]$, $[e^+,\bar\nu_e,e^+]$,
$[e^+,\bar\nu_\tau,e^+]$, $[\mu,e^+,e^+]$, $[\mu,\bar\nu_e,e^+]$,
$[\mu^+,\bar\nu_e,e^+]$,\\ $[t,e^+,e^+]$,
$[\tau^+,\bar\nu_e,e^+]$, $[\bar\nu_e,e^+,e^+]$,
$[\bar\nu_e,e^+,\mu]$, $[\bar\nu_e,e^+,\mu^+]$,
$[\bar\nu_e,e^+,\tau^+]$, $[\bar\nu_e,e^+,\nu_e]$,\\
$[\bar\nu_e,e^+,\nu_\mu]$, $[\bar\nu_e,e^+,\nu_\tau]$,
$[\bar\nu_e,\tau^+,\bar\nu_e]$, $[\bar\nu_e,\bar\nu_e,e^+]$,
$[\bar\nu_e,\bar\nu_e,\tau^+]$, $[\bar\nu_e,\bar\nu_e,\nu_\mu]$,
$[\bar\nu_e,\bar\nu_e,\nu_\tau]$,\\
$[\bar\nu_e,\nu_\mu,\bar\nu_e]$, $[\bar\nu_e,\nu_\tau,\bar\nu_e]$,
$[\bar\nu_\tau,e^+,e^+]$, $[\nu_e,\bar\nu_e,e^+]$,
$[\nu_\mu,\bar\nu_e,e^+]$, $[\nu_\tau,\bar\nu_e,e^+]$.}

\paragraph{Remark 5.5.} In general. the hyperoperation $" \otimes
"$ is not associative.\\

\paragraph{Theorem 5.6.} If $(L, \otimes)$ is above $H_v$-group,
then the following statements hold. {\it \begin{itemize}
\item[1)] There is not any
$H_v$-subgroups of order $5$, $7$, $8$, $9$, $10$, and, $11$ for
$(L, \otimes)$;
\item[2)] \medskip  All the $1-$dimensional $H_v$-subgroups of $L$ are:
\begin{eqnarray*}
\begin{array}{lclclclcl}
L^1_1=\{e\},&&L^1_2=\{e^+\},&&L^1_3=\{\mu\},&&L^1_4=\{\mu^+\},&&L^1_5=\{\tau\},\\
L^1_6=\{\tau^+\},&&L^1_7=\{\bar\nu_e\},&&L^1_8=\{\bar\nu_\mu\},&&L^1_9=\{\bar\nu_\tau\},&&L^1_{10}=
\{\nu_e\},\\
L^1_{11}=\{\nu_\mu\},&&L^1_{12}=\{\nu_\tau\}
\end{array}
\end{eqnarray*}
\item[3)] All the $2-$dimensional $H_v$-subgroups of $L$ are:
\begin{eqnarray*}
\begin{array}{lclclcl}
L^2_{1}=\{e,\mu\},&&L^2_{2}= \{e,\tau\},&&L^2_{3}=\{e,\bar\nu_\mu\},&&L^2_{4}= \{e,\bar\nu_\tau\},\\
L^2_{5}= \{e,\nu_e\},&&L^2_{6}= \{e^+,\mu^+\},&&L^2_{7}= \{e^+,\tau^+\},&&L^2_{8}= \{e^+,\nu_\mu\},\\
L^2_{9}= \{e^+,\nu_\tau\},&&L^2_{10}= \{\mu,\tau\},&&L^2_{11}= \{\mu,\bar\nu_e\},&&L^2_{12}= \{\mu,\bar\nu_\tau\},\\
L^2_{13}= \{\mu,\nu_\mu\},&&L^2_{14}=
\{\mu^+,\tau^+\},&&L^2_{15}=\{\mu^+,\bar\nu_\mu\},&&L^2_{16}=
 \{\mu^+,\nu_e\},\\
L^2_{17}=\{\mu^+,\nu_\tau\},&&L^2_{18}=\{\tau,\bar\nu_e\},&&L^2_{19}=\{\tau,\bar\nu_\mu\},&&L^2_{20}=
\{\tau,\nu_\tau\},\\
L^2_{21}= \{\tau^+,\bar\nu_\tau\},&&L^2_{22}= \{\tau^+,\nu_e\},&&L^2_{23}= \{\tau^+,\nu_\mu\},&&L^2_{24}=
 \{\bar\nu_e,\bar\nu_\mu\},\\
L^2_{25}=
\{\bar\nu_e,\bar\nu_\tau\},&&L^2_{26}=\{\bar\nu_\mu,\bar\nu_\tau\},&&L^2_{27}=
\{\nu_e,\nu_\mu\},&& L^2_{28}= \{\nu_e,\nu_\tau\},\\ L^2_{29}=
\{\nu_\mu,\nu_\tau\}.
\end{array}
\end{eqnarray*}
\item[4)] All the $3-$dimensional $H_v$-subgroups of $L$ are:
\begin{eqnarray*}
\begin{array}{lclcl}
L^3_1= \{ e,\mu,\tau \},&&L^3_2= \{ e,\mu,\bar\nu_\tau \},&&L^3_3= \{ e,\tau,\bar\nu_\mu \} ,\\
L^3_4= \{ e,\bar\nu_\mu,\bar\nu_\tau \},&&L^3_5= \{ e^+,\mu^+,\tau^+ \},&&L^3_6= \{ e^+,\mu^+,{\it \nu_\tau} \},\\
L^3_7= \{ e^+,\tau^+,\nu_\mu \},&&L^3_8= \{ e^+,\nu_\mu,\nu_e\},&&L^3_9= \{ \mu,\tau ,\bar\nu_e \},\\
L^3_{10}= \{ \mu,\bar\nu_e,\bar\nu_\tau \},&&L^3_{11}= \{
\mu^+,\tau^+,\nu_e \},&&L^3_{12}=
\{ \mu^+,\nu_e,{\it \nu_\tau} \},\\
L^3_{13}= \{ \tau,\bar\nu_e,\bar\nu_\mu \},&&L^3_{14}= \{
\tau^+,\nu_e,\nu_\mu \},&& L^3_{15}=
\{\bar\nu_e,\bar\nu_\mu,\bar\nu_\tau \},\\
 L^3_{16}= \{ \nu_e,\nu_\mu,{\it \nu_\tau} \}.
\end{array}
\end{eqnarray*}
\item[5)] All the $4-$dimensional $H_v$-subgroups of $L$ are:
\begin{eqnarray*}
\begin{array}{lclcl}
L^4_1= \{ e,\mu,\nu_e,\nu_\mu \},&& L^4_2= \{ e,\mu^+,
\bar\nu_\mu,\nu_e \},&&
L^4_3= \{ e,\tau,\nu_e,{\it \nu_\tau} \},\\
L^4_4= \{ e,\tau^+,\bar\nu_\tau,\nu_e \},&& L^4_5= \{
e^+,\mu^+,\bar\nu_e,\bar\nu_\mu \},&&
L^4_6= \{ \mu,\tau,\nu_\mu,{\it \nu_\tau} \},\\
L^4_7= \{ \mu,\tau^+,\bar\nu_\tau,\nu_\mu \},&& L^4_8= \{
\mu^+,\tau,\bar\nu_\mu,{\it \nu_\tau} \},&& L^4_9= \{
\mu^+,\tau^+,\bar\nu_\mu,\bar\nu_\tau \}.
\end{array}
\end{eqnarray*}
\item[6)] All the $6-$dimensional $H_v$-subgroups of $L$ are:
\begin{eqnarray*}
\begin{array}{lcl}
L^6_1= \{ e,\mu,\tau,\bar\nu_e,\bar\nu_\mu,\bar\nu_\tau \},&&
L^6_2= \{ e,\mu,\tau,\nu_e,\nu_\mu,{\it \nu_\tau} \},\\
L^6_3= \{ e,\mu,\tau^+,\bar\nu_\tau,\nu_e,\nu_\mu \},&&
L^6_4= \{ e,\mu^+,\tau,\bar\nu_\mu,\nu_e,{\it \nu_\tau} \},\\
L^6_5= \{ e,\mu^+,\tau^+,\bar\nu_\mu,\bar\nu_\tau,\nu_e \},&&
L^6_6= \{ e^+,\mu^+,\tau,\bar\nu_e, \bar\nu_\mu,{\it \nu_tau} \},\\
L^6_7= \{ e^+,\mu^+,\tau^+,\bar\nu_e,\bar\nu_\mu,\bar\nu_\tau
\},&& L^6_8= \{ e^+,\mu^+,\tau^+ ,\nu_e,\nu_\mu,{\it \nu_\tau} \}.
\end{array}
\end{eqnarray*}
\end{itemize}
}
 The result follows immediately from Theorem 5.6 and Definitions
3.1, 3.2.\\

\paragraph{Conclusion 5.7.}
{\it If $(L, \otimes)$ is above $H_v$-group, then the following
statements (the inclusion $H_v$-subgroups ) hold:
\begin{itemize}
\item[(i)] for $1$-dimensional:\\

$L^1_1\subset L^2_1$,$\;$  $L^1_1\subset L^2_2$,$\;$ $L^1_1\subset
L^2_3$,$\;$  $L^1_1\subset L^2_4$,$\;$  $L^1_1\subset L^2_5$,$\;$
$L^1_1\subset L^3_1$,$\;$  $L^1_1\subset L^3_2$,$\;$ $L^1_1\subset
L^3_3$,$\;$  $L^1_1\subset L^3_4$,$\;$  $L^1_1\subset L^4_1$,$\;$
$L^1_1\subset L^4_2$,$\;$  $L^1_1\subset L^4_3$,$\;$ $L^1_1\subset
L^4_4$,$\;$  $L^1_1\subset L^6_1$,$\;$  $L^1_1\subset L^6_2$,$\;$
$L^1_1\subset L^6_3$,$\;$  $L^1_1\subset L^6_4$,$\;$ $L^1_1\subset
L^6_5$,$\;$  $L^1_1\subset L^{12}$,$\;$  $L^1_2\subset L^2_6$,$\;$
$L^1_2\subset L^2_7$,$\;$  $L^1_2\subset L^2_8$,$\;$ $L^1_2\subset
L^2_9$,$\;$  $L^1_2\subset L^3_5$,$\;$  $L^1_2\subset L^3_6$,$\;$
$L^1_2\subset L^3_7$,$\;$  $L^1_2\subset L^3_8$,$\;$ $L^1_2\subset
L^4_5$,$\;$  $L^1_2\subset L^6_6$,$\;$  $L^1_2\subset L^6_7$,$\;$
$L^1_2\subset L^6_8$,$\;$  $L^1_2\subset L^{12}$,$\;$
$L^1_3\subset L^2_1$,$\;$  $L^1_3\subset L^2_{10}$,$\;$
$L^1_3\subset L^2_{11}$,$\;$  $L^1_3\subset L^2_{12}$,$\;$
$L^1_3\subset L^2_13$,$\;$  $L^1_3\subset L^3_1$,$\;$
$L^1_3\subset L^3_2$,$\;$  $L^1_3\subset L^3_9$,$\;$ $L^1_3\subset
L^3_{10}$,$\;$  $L^1_3\subset L^4_1$,$\;$  $L^1_3\subset
L^4_6$,$\;$ $L^1_3\subset L^4_7$,$\;$  $L^1_3\subset L^6_1$,$\;$
$L^1_3\subset L^6_2$,$\;$  $L^1_3\subset L^6_3$,$\;$ $L^1_3\subset
L^{12}$,$\;$  $L^1_4\subset L^2_6$,$\;$  $L^1_4\subset
L^2_{14}$,$\;$ $L^1_4\subset L^2_{15}$,$\;$  $L^1_4\subset
L^2_{16}$,$\;$ $L^1_4\subset L^2_{17}$,$\;$  $L^1_4\subset
L^3_5$,$\;$ $L^1_4\subset L^3_6$,$\;$  $L^1_4\subset
L^3_{11}$,$\;$ $L^1_4\subset L^3_{12}$,$\;$  $L^1_4\subset
L^4_2$,$\;$ $L^1_4\subset L^4_5$,$\;$  $L^1_4\subset L^4_8$,$\;$
$L^1_4\subset L^4_9$,$\;$  $L^1_4\subset L^6_4$,$\;$ $L^1_4\subset
L^6_5$,$\;$ $L^1_4\subset L^6_6$,$\;$  $L^1_4\subset L^6_7$,$\;$
$L^1_4\subset L^6_8$,$\;$  $L^1_4\subset L^{12}$,$\;$
$L^1_5\subset L^2_2$,$\;$ $L^1_5\subset L^2_{10}$,$\;$
$L^1_5\subset L^2_{18}$,$\;$ $L^1_5\subset L^2_{19}$,$\;$
$L^1_5\subset L^2_{20}$,$\;$ $L^1_5\subset L^3_1$,$\;$
$L^1_5\subset L^3_3$,$\;$ $L^1_5\subset L^3_9$,$\;$  $L^1_5\subset
L^3_{13}$,$\;$ $L^1_5\subset L^4_3$,$\;$ $L^1_5\subset L^4_6$,$\;$
$L^1_5\subset L^4_8$,$\;$ $L^1_5\subset L^6_1$,$\;$  $L^1_5\subset
L^6_2$,$\;$ $L^1_5\subset L^6_4$,$\;$ $L^1_5\subset L^6_6$,$\;$
$L^1_5\subset L^{12}$,$\;$ $L^1_6\subset L^2_7$,$\;$ $L^1_6\subset
L^2_{14}$,$\;$ $L^1_6\subset L^2_{21}$,$\;$ $L^1_6\subset
L^2_{22}$,$\;$ $L^1_6\subset L^2_{23}$,$\;$ $L^1_6\subset
L^3_5$,$\;$ $L^1_6\subset L^3_7$,$\;$  $L^1_6\subset
L^3_{11}$,$\;$ $L^1_6\subset L^3_{14}$,$\;$  $L^1_6\subset
L^4_4$,$\;$ $L^1_6\subset L^4_7$,$\;$ $L^1_6\subset L^4_9$,$\;$
$L^1_6\subset L^6_3$,$\;$  $L^1_6\subset L^6_5$,$\;$ $L^1_6\subset
L^6_7$,$\;$ $L^1_6\subset L^6_8$,$\;$ $L^1_6\subset L^{12}$,$\;$
$L^1_7\subset L^2_{11}$,$\;$ $L^1_7\subset L^2_{18}$,$\;$
$L^1_7\subset L^2_{24}$,$\;$ $L^1_7\subset L^2_{25}$,$\;$
$L^1_7\subset L^3_9$,$\;$ $L^1_7\subset L^3_{10}$,$\;$
$L^1_7\subset L^3_{13}$,$\;$ $L^1_7\subset L^3_{15}$,$\;$
$L^1_7\subset L^4_5$,$\;$ $L^1_7\subset L^6_1$,$\;$ $L^1_7\subset
L^6_6$,$\;$  $L^1_7\subset L^6_7$,$\;$ $L^1_7\subset L^{12}$,$\;$
$L^1_8\subset L^2_3$,$\;$ $L^1_8\subset L^2_{15}$,$\;$
$L^1_8\subset L^2_{19}$,$\;$ $L^1_8\subset L^2_{24}$,$\;$
$L^1_8\subset L^2_{26}$,$\;$ $L^1_8\subset L^3_3$,$\;$
$L^1_8\subset L^3_4$,$\;$  $L^1_8\subset L^3_{13}$,$\;$
$L^1_8\subset L^3_{15}$,$\;$ $L^1_8\subset L^4_2$,$\;$
$L^1_8\subset L^4_5$,$\;$  $L^1_8\subset L^4_8$,$\;$ $L^1_8\subset
L^4_9$,$\;$ $L^1_8\subset L^6_1$,$\;$ $L^1_8\subset L^6_4$,$\;$
$L^1_8\subset L^6_5$,$\;$  $L^1_8\subset L^6_6$,$\;$ $L^1_8\subset
L^6_7$,$\;$ $L^1_8\subset L^{12}$,$\;$ $L^1_9\subset L^2_4$,$\;$
$L^1_9\subset L^2_{12}$,$\;$ $L^1_9\subset L^2_{21}$,$\;$
$L^1_9\subset L^2_{25}$,$\;$ $L^1_9\subset L^2_{26}$,$\;$
$L^1_9\subset L^3_2$,$\;$ $L^1_9\subset L^3_4$,$\;$ $L^1_9\subset
L^3_{10}$,$\;$ $L^1_9\subset L^3_{15}$,$\;$  $L^1_9\subset
L^4_4$,$\;$ $L^1_9\subset L^4_7$,$\;$ $L^1_9\subset L^4_9$,$\;$
$L^1_9\subset L^6_1$,$\;$  $L^1_9\subset L^6_3$,$\;$ $L^1_9\subset
L^6_5$,$\;$ $L^1_9\subset L^6_7$,$\;$ $L^1_9\subset L^{12}$,$\;$
$L^1_{10}\subset L^2_5$,$\;$ $L^1_{10}\subset L^2_{16}$,$\;$
$L^1_{10}\subset L^2_{22}$,$\;$ $L^1_{10}\subset L^2_{27}$,$\;$
$L^1_{10}\subset L^2_{28}$,$\;$ $L^1_{10}\subset L^3_{11}$,$\;$
$L^1_{10}\subset L^3_{12}$,$\;$ $L^1_{10}\subset L^3_{14}$,$\;$
$L^1_{10}\subset L^3_{16}$,$\;$ $L^1_{10}\subset L^4_1$,$\;$
$L^1_{10}\subset L^4_2$,$\;$ $L^1_{10}\subset L^4_3$,$\;$
$L^1_{10}\subset L^4_4$,$\;$ $L^1_{10}\subset L^6_2$,$\;$
$L^1_{10}\subset L^6_3$,$\;$ $L^1_{10}\subset L^6_4$,$\;$
$L^1_{10}\subset L^6_5$,$\;$ $L^1_{10}\subset L^6_8$,$\;$
$L^1_{10}\subset L^{12}$,$\;$ $L^1_{11}\subset L^2_8$,$\;$
$L^1_{11}\subset L^2_{13}$,$\;$ $L^1_{11}\subset L^2_{23}$,$\;$
$L^1_{11}\subset L^2_{27}$,$\;$ $L^1_{11}\subset L^2_{29}$,$\;$
$L^1_{11}\subset L^3_7$,$\;$ $L^1_{11}\subset L^3_8$,$\;$
$L^1_{11}\subset L^3_{14}$,$\;$ $L^1_{11}\subset L^3_{16}$,$\;$
$L^1_{11}\subset L^4_1$,$\;$ $L^1_{11}\subset L^4_6$,$\;$
$L^1_{11}\subset L^4_7$,$\;$ $L^1_{11}\subset L^6_2$,$\;$
$L^1_{11}\subset L^6_3$,$\;$ $L^1_{11}\subset L^6_8$,$\;$
$L^1_{11}\subset L^{12}$,$\;$ $L^1_{12}\subset L^2_9$,$\;$
$L^1_{12}\subset L^2_{17}$,$\;$ $L^1_{12}\subset L^2_{20}$,$\;$
$L^1_{12}\subset L^2_{28}$,$\;$ $L^1_{12}\subset L^2_{29}$,$\;$
$L^1_{12}\subset L^3_6$,$\;$ $L^1_{12}\subset L^3_8$,$\;$
$L^1_{12}\subset L^3_{12}$,$\;$ $L^1_{12}\subset L^3_{16}$,$\;$
$L^1_{12}\subset L^4_3$,$\;$ $L^1_{12}\subset L^4_6$,$\;$
$L^1_{12}\subset L^4_8$,$\;$ $L^1_{12}\subset L^6_2$,$\;$
$L^1_{12}\subset L^6_4$,$\;$ $L^1_{12}\subset L^6_6$,$\;$
$L^1_{12}\subset L^6_8$,$\;$ $L^1_{12}\subset L^{12}$.

\item[(i)] for $2$-dimensional:\\

$L^2_1\subset L^3_1$,$\;$  $L^2_1\subset L^3_2$,$\;$ $L^2_1\subset
L^4_1$,$\;$  $L^2_1\subset L^6_1$,$\;$  $L^2_1\subset L^6_2$,$\;$
$L^2_1\subset L^6_3$,$\;$  $L^2_1\subset L^{12}$,$\;$
$L^2_2\subset L^3_1$,$\;$  $L^2_2\subset L^3_3$,$\;$ $L^2_2\subset
L^4_3$,$\;$  $L^2_2\subset L^6_1$,$\;$  $L^2_2\subset L^6_2$,$\;$
$L^2_2\subset L^6_4$,$\;$  $L^2_2\subset L^{12}$,$\;$
$L^2_3\subset L^3_3$,$\;$  $L^2_3\subset L^3_4$,$\;$ $L^2_3\subset
L^4_2$,$\;$  $L^2_3\subset L^6_1$,$\;$  $L^2_3\subset L^6_4$,$\;$
$L^2_3\subset L^6_5$,$\;$  $L^2_3\subset L^{12}$,$\;$
$L^2_4\subset L^3_2$,$\;$  $L^2_4\subset L^3_4$,$\;$ $L^2_4\subset
L^4_4$,$\;$  $L^2_4\subset L^6_1$,$\;$  $L^2_4\subset L^6_3$,$\;$
$L^2_4\subset L^6_5$,$\;$  $L^2_4\subset L^{12}$,$\;$
$L^2_5\subset L^4_1$,$\;$  $L^2_5\subset L^4_2$,$\;$ $L^2_5\subset
L^4_3$,$\;$  $L^2_5\subset L^4_4$,$\;$  $L^2_5\subset L^6_2$,$\;$
$L^2_5\subset L^6_3$,$\;$  $L^2_5\subset L^6_4$,$\;$ $L^2_5\subset
L^6_5$,$\;$  $L^2_5\subset L^{12}$,$\;$  $L^2_6\subset L^3_5$,$\;$
$L^2_6\subset L^3_6$,$\;$  $L^2_6\subset L^4_5$,$\;$ $L^2_6\subset
L^6_6$,$\;$  $L^2_6\subset L^6_7$,$\;$  $L^2_6\subset L^6_8$,$\;$
$L^2_6\subset L^{12}$,$\;$  $L^2_7\subset L^3_5$,$\;$
$L^2_7\subset L^3_7$,$\;$  $L^2_7\subset L^6_7$,$\;$ $L^2_7\subset
L^6_8$,$\;$  $L^2_7\subset L^{12}$,$\;$  $L^2_8\subset L^3_7$,$\;$
$L^2_8\subset L^3_8$,$\;$  $L^2_8\subset L^6_8$,$\;$ $L^2_8\subset
L^{12}$,$\;$  $L^2_9\subset L^3_6$,$\;$  $L^2_9\subset L^3_8$,$\;$
$L^2_9\subset L^6_6$,$\;$  $L^2_9\subset L^6_8$,$\;$ $L^2_9\subset
L^{12}$,$\;$  $L^2_{10}\subset L^3_1$,$\;$  $L^2_{10}\subset
L^3_9$,$\;$  $L^2_{10}\subset L^4_6$,$\;$  $L^2_{10}\subset
L^6_1$,$\;$  $L^2_{10}\subset L^6_2$,$\;$  $L^2_{10}\subset
L^{12}$,$\;$  $L^2_{11}\subset L^3_9$,$\;$  $L^2_{11}\subset
L^3_{10}$,$\;$  $L^2_{11}\subset L^6_1$,$\;$  $L^2_{11}\subset
L^{12}$,$\;$  $L^2_{12}\subset L^3_2$,$\;$  $L^2_{12}\subset
L^3_{10}$,$\;$  $L^2_{12}\subset L^4_7$,$\;$  $L^2_{12}\subset
L^6_1$,$\;$  $L^2_{12}\subset L^6_3$,$\;$  $L^2_{12}\subset
L^{12}$,$\;$  $L^2_{13}\subset L^4_1$,$\;$  $L^2_{13}\subset
L^4_6$,$\;$ $L^2_{13}\subset L^4_7$,$\;$  $L^2_{13}\subset
L^6_2$,$\;$ $L^2_{13}\subset L^6_3$,$\;$  $L^2_{13}\subset
L^{12}$,$\;$ $L^2_{14}\subset L^3_5$,$\;$  $L^2_{14}\subset
L^3_{11}$,$\;$ $L^2_{14}\subset L^4_9$,$\;$  $L^2_{14}\subset
L^6_5$,$\;$ $L^2_{14}\subset L^6_7$,$\;$  $L^2_{14}\subset
L^6_8$,$\;$ $L^2_{14}\subset L^{12}$,$\;$  $L^2_{15}\subset
L^4_2$,$\;$ $L^2_{15}\subset L^4_5$,$\;$  $L^2_{15}\subset
L^4_8$,$\;$ $L^2_{15}\subset L^4_9$,$\;$  $L^2_{15}\subset
L^6_4$,$\;$ $L^2_{15}\subset L^6_5$,$\;$  $L^2_{15}\subset
L^6_6$,$\;$ $L^2_{15}\subset L^6_7$,$\;$  $L^2_{15}\subset
L^{12}$,$\;$ $L^2_{16}\subset L^3_{11}$,$\;$  $L^2_{16}\subset
L^3_{12}$,$\;$ $L^2_{16}\subset L^4_2$,$\;$  $L^2_{16}\subset
L^6_4$,$\;$ $L^2_{16}\subset L^6_5$,$\;$  $L^2_{16}\subset
L^6_8$,$\;$ $L^2_{16}\subset L^{12}$,$\;$  $L^2_{17}\subset
L^3_6$,$\;$ $L^2_{17}\subset L^3_{12}$,$\;$  $L^2_{17}\subset
L^4_8$,$\;$ $L^2_{17}\subset L^6_4$,$\;$  $L^2_{17}\subset
L^6_6$,$\;$ $L^2_{17}\subset L^6_8$,$\;$  $L^2_{17}\subset
L^{12}$,$\;$ $L^2_{18}\subset L^3_9$,$\;$  $L^2_{18}\subset
L^3_{13}$,$\;$ $L^2_{18}\subset L^6_1$,$\;$  $L^2_{18}\subset
L^6_6$,$\;$ $L^2_{18}\subset L^{12}$,$\;$  $L^2_{19}\subset
L^3_3$,$\;$ $L^2_{19}\subset L^3_{13}$,$\;$  $L^2_{19}\subset
L^4_8$,$\;$ $L^2_{19}\subset L^6_1$,$\;$  $L^2_{19}\subset
L^6_4$,$\;$ $L^2_{19}\subset L^6_6$,$\;$  $L^2_{19}\subset
L^{12}$,$\;$ $L^2_{20}\subset L^4_3$,$\;$  $L^2_{20}\subset
L^4_6$,$\;$ $L^2_{20}\subset L^4_8$,$\;$  $L^2_{20}\subset
L^6_2$,$\;$ $L^2_{20}\subset L^6_4$,$\;$  $L^2_{20}\subset
L^6_6$,$\;$ $L^2_{20}\subset L^{12}$,$\;$  $L^2_{21}\subset
L^4_4$,$\;$ $L^2_{21}\subset L^4_7$,$\;$  $L^2_{21}\subset
L^4_9$,$\;$ $L^2_{21}\subset L^6_3$,$\;$  $L^2_{21}\subset
L^6_5$,$\;$ $L^2_{21}\subset L^6_7$,$\;$  $L^2_{21}\subset
L^{12}$,$\;$ $L^2_{22}\subset L^3_{11}$,$\;$  $L^2_{22}\subset
L^3_{14}$,$\;$ $L^2_{22}\subset L^4_4$,$\;$  $L^2_{22}\subset
L^6_3$,$\;$ $L^2_{22}\subset L^6_5$,$\;$  $L^2_{22}\subset
L^6_8$,$\;$ $L^2_{22}\subset L^{12}$,$\;$  $L^2_{23}\subset
L^3_7$,$\;$ $L^2_{23}\subset L^3_{14}$,$\;$  $L^2_{23}\subset
L^4_7$,$\;$ $L^2_{23}\subset L^6_3$,$\;$  $L^2_{23}\subset
L^6_8$,$\;$ $L^2_{23}\subset L^{12}$,$\;$  $L^2_{24}\subset
L^3_{13}$,$\;$ $L^2_{24}\subset L^3_{15}$,$\;$  $L^2_{24}\subset
L^4_5$,$\;$ $L^2_{24}\subset L^6_1$,$\;$  $L^2_{24}\subset
L^6_6$,$\;$ $L^2_{24}\subset L^6_7$,$\;$  $L^2_{24}\subset
L^{12}$,$\;$ $L^2_{25}\subset L^3_{10}$,$\;$  $L^2_{25}\subset
L^3_{15}$,$\;$ $L^2_{25}\subset L^6_1$,$\;$  $L^2_{25}\subset
L^6_7$,$\;$ $L^2_{25}\subset L^{12}$,$\;$  $L^2_{26}\subset
L^3_4$,$\;$ $L^2_{26}\subset L^3_{15}$,$\;$  $L^2_{26}\subset
L^4_9$,$\;$ $L^2_{26}\subset L^6_1$,$\;$  $L^2_{26}\subset
L^6_5$,$\;$ $L^2_{26}\subset L^6_7$,$\;$  $L^2_{26}\subset
L^{12}$,$\;$ $L^2_{27}\subset L^3_{14}$,$\;$  $L^2_{27}\subset
L^3_{16}$,$\;$ $L^2_{27}\subset L^4_1$,$\;$  $L^2_{27}\subset
L^6_2$,$\;$ $L^2_{27}\subset L^6_3$,$\;$  $L^2_{27}\subset
L^6_8$,$\;$ $L^2_{27}\subset L^{12}$,$\;$  $L^2_{28}\subset
L^3_{12}$,$\;$ $L^2_{28}\subset L^3_{16}$,$\;$  $L^2_{28}\subset
L^4_3$,$\;$ $L^2_{28}\subset L^6_2$,$\;$  $L^2_{28}\subset
L^6_4$,$\;$ $L^2_{28}\subset L^6_8$,$\;$  $L^2_{28}\subset
L^{12}$,$\;$ $L^2_{29}\subset L^3_8$,$\;$  $L^2_{29}\subset
L^3_{16}$,$\;$ $L^2_{29}\subset L^4_6$,$\;$  $L^2_{29}\subset
L^6_2$,$\;$ $L^2_{29}\subset L^6_8$,$\;$  $L^2_{29}\subset
L^{12}$.

\item[(i)] for $3$-dimensional:\\

$L^3_1\subset L^6_1$,$\;$  $L^3_1\subset L^6_2$,$\;$ $L^3_1\subset
L^{12}$,$\;$  $L^3_2\subset L^6_1$,$\;$  $L^3_2\subset L^6_3$,$\;$
$L^3_2\subset L^{12}$,$\;$  $L^3_3\subset L^6_1$,$\;$
$L^3_3\subset L^6_4$,$\;$  $L^3_3\subset L^{12}$,$\;$
$L^3_4\subset L^6_1$,$\;$  $L^3_4\subset L^6_5$,$\;$ $L^3_4\subset
L^{12}$,$\;$  $L^3_5\subset L^6_7$,$\;$  $L^3_5\subset L^6_8$,$\;$
$L^3_5\subset L^{12}$,$\;$  $L^3_6\subset L^6_6$,$\;$
$L^3_6\subset L^6_8$,$\;$  $L^3_6\subset L^{12}$,$\;$
$L^3_7\subset L^6_8$,$\;$  $L^3_7\subset L^{12}$,$\;$
$L^3_8\subset L^6_8$,$\;$  $L^3_8\subset L^{12}$,$\;$
$L^3_9\subset L^6_1$,$\;$  $L^3_9\subset L^{12}$,$\;$
$L^3_{10}\subset L^6_1$,$\;$  $L^3_{10}\subset L^{12}$,$\;$
$L^3_{11}\subset L^6_5$,$\;$  $L^3_{11}\subset L^6_8$,$\;$
$L^3_{11}\subset L^{12}$,$\;$  $L^3_{12}\subset L^6_4$,$\;$
$L^3_{12}\subset L^6_8$,$\;$  $L^3_{12}\subset L^{12}$,$\;$
$L^3_{13}\subset L^6_1$,$\;$  $L^3_{13}\subset L^6_6$,$\;$
$L^3_{13}\subset L^{12}$,$\;$  $L^3_{14}\subset L^6_3$,$\;$
$L^3_{14}\subset L^6_8$,$\;$  $L^3_{14}\subset L^{12}$,$\;$
$L^3_{15}\subset L^6_1$,$\;$  $L^3_{15}\subset L^6_7$,$\;$
$L^3_{15}\subset L^{12}$,$\;$  $L^3_{16}\subset L^6_2$,$\;$
$L^3_{16}\subset L^6_8$,$\;$  $L^3_{16}\subset L^{12}$.

\item[(i)] for $4$-dimensional:\\

$L^4_1\subset L^6_2$,$\;$  $L^4_1\subset L^6_3$,$\;$ $L^4_1\subset
L^{12}$,$\;$  $L^4_2\subset L^6_4$,$\;$  $L^4_2\subset L^6_5$,$\;$
$L^4_2\subset L^{12}$,$\;$  $L^4_3\subset L^6_2$,$\;$
$L^4_3\subset L^6_4$,$\;$  $L^4_3\subset L^{12}$,$\;$
$L^4_4\subset L^6_3$,$\;$  $L^4_4\subset L^6_5$,$\;$ $L^4_4\subset
L^{12}$,$\;$  $L^4_5\subset L^6_6$,$\;$  $L^4_5\subset L^6_7$,$\;$
$L^4_5\subset L^{12}$,$\;$  $L^4_6\subset L^6_2$,$\;$
$L^4_6\subset L^{12}$,$\;$  $L^4_7\subset L^6_3$,$\;$
$L^4_7\subset L^{12}$,$\;$  $L^4_8\subset L^6_4$,$\;$
$L^4_8\subset L^6_6$,$\;$  $L^4_8\subset L^{12}$,$\;$
$L^4_9\subset L^6_5$,$\;$  $L^4_9\subset L^6_7$,$\;$
$L^4_9\subset L^{12}$.

\item[(i)] for $6$-dimensional:\\

$L^6_1\subset L^{12}$,$\;$  $L^6_2\subset L^{12}$,$\;$
$L^6_3\subset L^{12}$,$\;$  $L^6_4\subset L^{12}$,$\;$
$L^6_5\subset L^{12}$,$\;$  $L^6_6\subset L^{12}$,$\;$
$L^6_7\subset L^{12}$,$\;$  $L^6_8\subset L^{12}$.
\end{itemize}}

\end{document}